# The unified ballooning theory with weak up-down asymmetric mode structure and the numerical studies

T.Xie[1,2,*], H.Qin[1,3], Y.Z.Zhang[4], S.M.Mahajan[5]

[1]*Department of Modern Physics, University of Science and Technology of China, Hefei, Anhui 230026, China*

[2] *Key Laboratory of Geospace Environment, CAS, Hefei, Anhui 230026, China*

[3]*Plasma Physics Laboratory, Princeton University, P.O. Box 451, Princeton, New Jersey 08543, USA*

[4]*Center for Magnetic Fusion Theory, CAS, Hefei, Anhui 230026, China*

[5]*Institute for Fusion Studies, University of Texas at Austin, Austin, Texas 78712, USA*

**Abstract**

A unified ballooning theory, constructed on the basis of two special theories [Y. Z. Zhang, S. M. Mahajan, X. D. Zhang, Phys. Fluids B**4**, 2729 (1992); Y. Z. Zhang, T. Xie, Nucl. Fusion & Plasma Phys. **33**, 193 (2013)], shows that a *weak up-down asymmetric* mode structure is *normally* formed in an *up-down symmetric equilibrium*; the weak up-down asymmetry in mode structure is the manifestation of non-trivial higher order effects beyond the standard ballooning equation. It is shown that the asymmetric mode may have even higher growth rate than symmetric modes. Salient features of the theory are illustrated by investigating a fluid model for the ion temperature gradient (ITG) mode. The two dimensional (2D) analytical form of ITG mode, solved in ballooning representation, is then converted into the radial-poloidal space to provide the natural boundary condition for solving the 2D mathematical local eigenmode problem. We find the analytical expression of mode structure in good agreement with finite difference solution. This sets a reliable framework for quasi-linear computation.

## 1  Introduction

The ballooning theory is the asymptotic theory for the 2D local linear eigenmode problem dealing with high toroidal mode number ($n$) micro-turbulence in axisymmetric toroidal systems with a sheared magnetic field. Translational invariance to the lowest order of $1/n$, allows the dominant features of the 2D system to be captured by a one-dimensional (1D) differential equation, the so-called ballooning equation. Beyond this lowest order, all higher order equations contain the contribution from all translational symmetry breaking (TSB) terms. And if the higher-order TSB terms are not negligible, the full eigenmode problem would be moderately complicated.

We attempt, here, to present a more comprehensive 2D theory to explore the effects of TSB. The essence of the new aspects of the theory will be demonstrated by assuming a conventional up-down symmetric equilibrium. *The principal qualitatively new result of this detailed 2D ballooning theory is that the higher order TSB terms induce a weak up-down asymmetry in mode structure even in an up-down symmetric equilibrium.*

We begin with a somewhat comprehensive history of the linear mode theory based on the ballooning transform. The framework for the ID ballooning theory was constructed in the pioneering work [1-7] of Connor-Hastie-Taylor [1,5,6], and of Lee-Van Dam, Glasser, Pegoraro-Schep [2-4,7]. The 2D extension of Connor-Hastie-Taylor was carried out by Dewar [8,9], while Zhang-Mahajan (ZM) [10,11] built their 2D theories following the Lee-Van Dam, Glasser, Pegoraro-Schep lead.

From late 1970s to early 1990s, the 1D ballooning program to be called, collectively, as conventional ballooning theory (CBT), was applied very widely and with great success to micro-turbulence[12], toroidal drift waves[13-17], linear and nonlinear gyro-kinetics[18-22], ideal and resistive ballooning modes[23,24], and many other phenomena too numerate to mention. There was a shared near-consensus during this period that the 2D eigenvalue problem can, in essence, be reduced to solving the 1D ballooning equation with only one parameter named $\eta_0$ in [1,5,6] and $\lambda$ in [2-4,7]; this parameter was to be either $0$ or $\pi$ as determined from the so-called solvability condition $\partial \Omega(\eta_0)/\partial \eta_0 = 0$, where $\Omega(\eta_0)$ is the 'local' eigenvalue of the 'Schrödinger-like' ballooning equation at $\eta_0$, a quantity that pins the mode localization in poloidal angle ($\vartheta$).

Invoking the solvability condition has its root in the requirement for asymptotic convergence. In fact, the true expansion parameter of CBT is not $1/n$, but $1/\sqrt{n}$ constructed as the product of $1/n$ with the rate of variation in mode amplitude. The rate of variation, thus, has to be much slower than $n$, otherwise the asymptotic expansion could not be, justifiably, truncated. The variation rate consistent with the solvability condition is found order of $\sqrt{n}$. When the solvability condition leads to an $\eta_0$ (either $0$ or $\pi$) independent of higher order terms, the TSB does not contribute importantly to the final result; the TSB correction to eigenvalue ($\sim 1/n$) is negligible. The parameter $\eta_0$ corresponds, in the 2D ZM representation (also known as Fourier-ballooning representation) to $\lambda_0$, around which the poloidal localization is determined; in the 2D extension, the phase parameter $\lambda$ in Lee-Van Dam representation [2] becomes the continuous variable signifying the second dimension.

The emergence of $1/\sqrt{n}$ ordering has been shown (in the higher order theory) in CBT representation for some appropriate models [5,6]. In the Fourier-ballooning representation, this is equivalent to combined parity (CP) conservation in the 2D ballooning solution; the CP conservation is the invariance under the combined reflection $(k,\lambda) \to (-k,-\lambda)$ [10,25], where $k$ and $\lambda$ label the two independent variables of the 2D (Fourier-ballooning) representation (see Eq. (1)). The ZM solvability condition, however, appears very differently from that of CBT, because in the higher order theory of Connor-Hastie-Taylor, the second dimension is the radial variable; In contrast, the second dimension of ZM theory is $\lambda$, and the ZM solvability condition is obtained by eliminating the first derivative term in the equation (see Eq. (3)) *i.e.*, $\bar{L}_1 = 0$ [10].

There are two remarkable features of the conventional CBT higher order theory: 1) perfect up-down symmetry in mode structure. Let us call it horizontal (HBT) to distinguish it from the vertical ballooning theory as that will emerge in the Fourier-ballooning representation, 2) The solvability condition for CBT (HBT) can be satisfied only at some solitary radial ($r$ or $x$) position(s) [6,26-28] restricting the solution to isolated regions in parameter space, namely, not encompassing much of radial domain [10]. This last feature, the "solitariness" of the solution is, surely, a cause for concern.

The initial challenge to the conventional CBT approach was, to some extent, triggered by the studies of the toroidal Alfven eigenmode (TAE) in early 1990s. The ballooning approach dealing with TAE was pioneered by Zonca-Chen[29,30] based on the 1D ballooning representation, then studied by Dewar-Zhang[31] and Zhang-Zhang-Mahajan[32,33]; the latter two were based on the 2D ballooning representation. The non-ballooning 2D approach to solving TAE was devised by Rosenbluth-Berk-Van Dam-Linberg [34,35].

The investigation of TAE via CBT runs into a serious problem because, near marginal stability, the TAE poloidal structure is essentially oscillatory rather than localized; consequently, the asymptotic convergence is not guaranteed for small $1/n$. To insure asymptotic convergence, new theories using a second small parameter (*e.g.*, inverse of large aspect ratio) were developed [29-32]. In a sense the solvability condition becomes irrelevant to the ballooning theory dealing with TAE. Such an idea was immediately adopted by Taylor-Connor-Wilson [26,27] to drift waves in weak toroidal coupling. The interest in this type of ballooning structure was then motivated by, deliberately, avoiding the solvability condition required by CBT; the traditional CBT was believed to be less significant to anomalous transport because of the "solitariness" of the solution [26-28].

Similar to TAE, the second small parameter for drift wave, weak toroidal coupling, makes this type of toroidal drift wave less localized in poloidal direction, and is likely to have lower growth rate. An alternative solution for drift waves, without resorting to weak toroidal coupling for the second small parameter, was later obtained for which the poloidal mode structure was sufficiently localized due to a finite growth rate [25, 36-39]. The second small parameter, therefore, seems to be spontaneously generated by the instability of drift waves. It may be appropriate to call such a mode as 'spontaneous up-down symmetry breaking drift wave', and the underlying theory could be termed as vertical ballooning theory (VBT) for more general scenarios, because the poloidal structure, now, is localized at either of $\pm\pi/2$, that is, it displays strong up-down asymmetry. By the same token the non-localized ballooning theory (for TAE or "weak toroidal-coupling" drift wave) can be figuratively called flat ballooning theory (FBT). VBT/FBT is also known as *general mode* in Ref. [37], where the CBT (HBT) is named to be *isolated mode*.
.
The strong up-down asymmetric structure has been observed in experiments [40] and recent numerical simulations [41,42]. Technically, in terms of the ZM higher order equation (see Eq. (3) below) only the second derivative term is retained in CBT (or HBT); the convergence of asymptotic expansion is acquired by eliminating the first derivative term in real $\lambda$ space. Alternatively, one could remove the second derivative term and seek for a second small parameter to enforce asymptotic convergence; a procedure of this kind was adopted for VBT and FBT.

However, neither HBT nor VBT/FBT could be the optimal theory of micro-turbulence significant to anomalous transport. One may speculate a theory somewhat between the two: a theory that predicts higher growth rate than VBT/FBT and is not so much plagued by the 'stringent' solvability condition as required by HBT. Fortunately, two such optimal theories do exist in literature [11,43]. Because asymmetry is the defining characteristic, it will be appropriate to coin a new name for this class -- weak asymmetric ballooning theory (WABT). The WABT is optimal because its localization around bad curvature region implies largest excitation rates, and flexible deviation from the perfect horizontal orientation saves it from the stringent constraint of the 'solitariness' alluded to earlier. As expected, the small deviation away from equatorial plane is determined by all TBS terms. In a sense WABT is the manifestation of non-trivial higher order effects in the ballooning theory.

The two existing versions of WABT can be viewed as extensions of CBT reconstructed in 2D (Fourier-ballooning) representation (HBT), albeit, in different approaches. In the 1992-theory (WABT-92) [11] the solvability condition $\bar{L}_1 = 0$ is modified to $\mathcal{F} \equiv \bar{L}_1 + \Delta(\lambda_I) = 0$, where $\lambda_I \equiv \mathrm{Im}\,\lambda$, and $\Delta(\lambda_I)$, with $\Delta(\lambda_I = 0) = 0$, is the part of fast variation contributed from $\Omega(\lambda)$ which is, simultaneously, the potential in the equation in the second dimension, and the 'local' eigenvalue of the ballooning equation with (finite) $\lambda_I$. The adjustable $\lambda_I$ cures the abovementioned (no solution or solitariness) shortcomings of the HBT as shown in the example of Ref.[11], *i.e.*, $\mathcal{F} = 0$ was shown to be satisfied in a continuous radial section . The finite $\lambda_I$ breaks the CP conservation of the 2D wave function, and yields the weak up-down asymmetry.

In the 2013 version of the theory (WABT-13) [43], a new small (complex) parameter $\Xi \equiv \bar{L}_1 / \bar{L}_2$ is introduced to replace the rigorous solvability condition $\bar{L}_1 = 0$ for purely real $\lambda$ in HBT; also weak up-down asymmetry follows. This $\Xi$ plays the role of the second small parameter for the asymptotic convergence as the solvability condition $\bar{L}_1 = 0$ is *not* rigorously satisfied. These two existing special versions of WABT can be combined into one unified theory under the framework of WABT-13 with generally non-zero $\lambda_I$, by introducing $\Xi_1 \equiv \mathcal{F} / \bar{L}_2$ to be the second small parameter. Therefore, WABT-92 is just the limit of $\Xi_1 \equiv \mathcal{F}/\bar{L}_2 \to 0$ with $\lambda_I$ being the solution of $\mathcal{F} = 0$.

The unified WABT is described in some detail in section 2, and applied to the fluid ITG model in section 3. The basic equations of the model are derived in sub-section 3.1 first for real space ($(x,\vartheta)$ also discrete-poloidal $(x,l)$) representation; then in 2D $(k,\lambda)$ ballooning representation including contributions from the higher order terms. We solve the ITG model for WABT using a numerical shooting code; the results are presented graphically in sub-section 3.2, and are compared with those of existing theories, HBT, VBT. In sub-section 3.3, the parameterized analytical mode structure is derived and converted into the real space ($(x,l)$ also $(x,\vartheta)$) representation. The analytical expression serves to provide the natural boundary condition for the mathematical 2D partial differential equation. *It is, perhaps, the first time that the 2D asymptotic theory has been invoked to provide natural boundary condition to, numerically, solve local eigenmodes problem for tokamak physics.* In section 4 the finite difference method is adopted to solve the ITG system in real space $(x,l)$ representation by taking WABT to be the natural boundary condition using the analytical form; the numerical results and also compared with the analytical form of WABT throughout the entire region. Good agreements in the analytical and numerical work are found not only in eigenvalue but also in mode structure. Major results of the paper are summarized and discussed in section 5.

## 2 The unified WABT

This section begins with the introduction of higher order ballooning theory in Fourier-ballooning representation. From this system, the CBT in the Connor-Hastie-Taylor representation is, first, reconstructed (HBT), and then the extension to WABT is developed.

The equilibrium of large aspect ratio tokamak is assumed up-down symmetric, described by toroidal coordinates $(r,\vartheta,\varsigma)$, where $r$ measures radial position in minor radius, $\vartheta$ is the poloidal angle, $\varsigma$ is the toroidal angle. Owing to the axisymmetry, the 2D wave function $\varphi_l(x)$ for a mode with high toroidal mode number ($n$) (localized at the rational surface $r_0$ in safety factor $q(r_0)$ and magnetic shear $\hat{s}$), can be expressed in terms of two variables: the continuous radial variable $x \equiv nq(r_0)\hat{s}(r-r_0)/r_0$, and the discrete poloidal variable $l$, that labels the sideband coupled to the central Fourier mode, ($m \equiv nq(r_0)$), i.e. $\varphi_n(r,\vartheta,\varsigma) \equiv \exp(in\varsigma)\varphi_n(r,\vartheta)$, $\varphi_n(r,\vartheta) \equiv \exp(-im\vartheta)\sum_l \varphi_l(x)\exp(-il\vartheta)$.

The 2D ballooning transformation [10,11]

$$\varphi_l(x) = \frac{1}{2\pi}\int_{-\pi}^{\pi} d\lambda \int_{-\infty}^{+\infty} dk\, e^{ik(x-l)-i\lambda l}\varphi(k,\lambda) \tag{1}$$

defines the wave function $\varphi(k,\lambda)$ in the Fourier space. It obeys

$$\left[L_0 + \frac{iL_1}{n}\frac{\partial}{\partial\lambda} + \frac{L_2}{n^2}\frac{\partial^2}{\partial\lambda^2} + \cdots - \Omega\right]\varphi(k,\lambda) = 0, \tag{2}$$

where $L_0 \equiv L_0[k,\partial/\partial k;\lambda]$ is the ballooning operator, and $L_1$, $L_2$ are also periodic functions of $\lambda$. The attempt to solve this system begins with first "solving" the ballooning differential equation $[L_0 - \Omega(\lambda)]\chi(k,\lambda) = 0$ for the $\lambda$-parametrized local eigen-value $\Omega(\lambda)$. This $\Omega(\lambda)$ is then fed into Eq. (2), whose eigen-solutions will yield the global eigen-value $\Omega$ and the corresponding wave function. After the wave function is factorized to be $\varphi(k,\lambda) = \psi(\lambda)\chi(k,\lambda)$, the second step consists of solving the ordinary differential equation in the second dimension,

$$\left(\frac{i\bar{L}_1}{n}\frac{d}{d\lambda} + \frac{\bar{L}_2}{n^2}\frac{d^2}{d\lambda^2} + \cdots - [\Omega - \Omega(\lambda)]\right)\psi(\lambda) = 0, \tag{3}$$

where $\bar{L}_\alpha$ ($\alpha = 1,2,\cdots$) are averages of the operators $L_\alpha$ defined below in Eqs. (4)-(6).

Both the HBT (CBT in Fourier-ballooning representation) and the WABT are asymptotic theories based on Eq. (3) that has been truncated at the second derivative term. The truncation yields faster variation of $\psi(\lambda)$ than $\chi(k,\lambda)$ in $\lambda$ as long as $\partial\ln\psi(\lambda)/\partial\lambda \gg 1$ as shown below. It reads like

$$\frac{d^2\psi}{d\lambda^2} + P(\lambda)\frac{d\psi}{d\lambda} + Q(\lambda)\psi(\lambda) = 0, \tag{4}$$

where

$$P(\lambda) \equiv \frac{n}{\bar{L}_2^{(0)}}\left(i\bar{L}_1^{(0)} + \frac{2\bar{L}_2^{(1)}}{n}\right), \quad Q(\lambda) \equiv \frac{n^2}{\bar{L}_2^{(0)}}\left(\Omega(\lambda) - \Omega + \frac{i\bar{L}_1^{(1)}}{n} + \frac{\bar{L}_2^{(2)}}{n^2}\right), \tag{5}$$

$$\bar{L}_\alpha^{(j)} \equiv \int_{-\infty}^{\infty} dk\, \chi^* L_\alpha \frac{\partial^j \chi}{\partial\lambda^j} \Big/ \int_{-\infty}^{\infty} dk\, \chi^*\chi, \quad (j=0,1,2;\ \alpha=1,2), \tag{6}$$

where $\chi^*$ is the complex conjugate of $\chi$. The asymptotic convergence of Eq. (3) with truncation requires that the condition $(1/n)(\partial \ln \psi(\lambda)/\partial \lambda) \ll 1$ be satisfied.

The HBT (CBT in Fourier-ballooning representation) is the consistent solution obtained by resorting to the solvability condition $\bar{L}_1 = 0$, while satisfying the CP conservation, *i.e.*, the wave function $\varphi(k,\lambda)$ is invariant under the combined reflection $(k,\lambda) \to (-k,-\lambda)$ [10,25]. However, the CP conservation requires that $\lambda$ be real, and the condition $\bar{L}_1 = 0$ may even not, in general, be satisfied at any point in the entire radial domain [10] save for specially designed *ad hoc* models [5,6]. On analytic continuation to the complex plane, $\lambda \to \lambda_r + i\lambda_I$, the CP conservation is violated, which leads to the WABT-92[11], for which the HBT solvability condition $\bar{L}_1 = 0$ is replaced by the new solvability condition $\mathcal{F} \equiv \bar{L}_1 + \Delta(\lambda_I) = 0$. As shown in Ref. [11], for drift waves, this condition can be satisfied in a certain continuous domain of parameter space. For both HBT and WABT-92 the asymptotic convergence is determined by $\partial \ln \psi(\lambda)/\partial \lambda \sim \sqrt{n}$. However, WABT-92 is not the only existing theory of WABT.

The new WABT, namely WABT-13, is the alternative approach to seek for solutions of Eq. (4) satisfying the fundamental condition for the validity of asymptotic expansion, $(1/n)(d/d\lambda) \ll 1$ with real $\lambda$.

The substitution

$$\psi(\lambda) = \Phi(\lambda) \exp\left[-\frac{1}{2}\int^{\lambda} d\lambda' P(\lambda')\right] \tag{7}$$

eliminates the first derivative in the equation for $\Phi(\lambda)$

$$\frac{d^2\Phi}{d\lambda^2} + \left[Q(\lambda) - \frac{P^2}{4} - \frac{1}{2}\frac{dP}{d\lambda}\right]\Phi = 0. \tag{8}$$

Equation (8) is similar to the HBT equation in the second dimension equation with $d \ln \Phi / d\lambda \sim \sqrt{n}$. However, both $\psi(\lambda)$ and the global eigen-value can, and will differ from the predictions of the HBT. The salient feature of WABT-13 can be readily seen by making the following simplifications, (a) letting $\bar{L}_\alpha \approx \bar{L}_\alpha^{(0)}$ by neglecting higher order terms such as $\bar{L}_\alpha^{(j)}$ ($j > 0$) and $dP/d\lambda$, (b) neglecting the weakly $\lambda$-dependent part in $\bar{L}_\alpha^{(0)}$, and (c) approximating $\Omega(\lambda)$ by $\Omega_0 + \Omega_1 \cos\lambda \approx \Omega_0 + \Omega_1(1-\lambda^2/2)$. Equation (8), then, reduces to the well-known

Weber equation whose fundamental solution is $\Phi(\lambda) \approx \exp\left[-\sqrt{\Omega_1/2\bar{L}_2}\, n\lambda^2/2\right]$ with the global eigen-value

$$\Omega = \Omega_0 + \Omega_1 + \frac{(\bar{L}_1)^2}{4\bar{L}_2}. \tag{9}$$

The full wave function may be constructed as

$$\psi(\lambda) = \Phi(\lambda)\exp\left(-in\frac{\bar{L}_1}{2\bar{L}_2}\lambda\right). \tag{10}$$

To satisfy $d\ln\psi/d\lambda \ll n$, the constraint necessary for the validity of the asymptotic expansion, one must demand $|\Xi| \equiv |\bar{L}_1/\bar{L}_2| \ll 1$. Here, $\Xi$ plays the role of second small parameter of WABT. Notice that the modification induced by $\bar{L}_1$ to the global eigen-value in Eq. (9), $(\bar{L}_1)^2/4\bar{L}_2 = \Xi\bar{L}_1/4$, can be arbitrary even for $|\Xi| \ll 1$. The exponential factor induced by $\bar{L}_1$ in Eq. (10) breaks the $\lambda$-inversion symmetry, and thus the up-down symmetry. The strong localization of the new WABT eigen-function at $\lambda = 0$ ($\Phi(\lambda) \approx \exp\left[-\sqrt{\Omega_1/2\bar{L}_2}\, n\lambda^2/2\right]$) distinguishes it decisively from the mode structure in the VBT.

The unified WABT is constructed within the framework of WABT-13, but allowing complex $\lambda$. To the leading order Eq. (8) can be cast into the form

$$\frac{d^2\Phi}{d\lambda^2} + \frac{n^2}{\bar{L}_2}\left[\frac{\bar{L}_1^2}{4\bar{L}_2} - (\Omega - \Omega(\lambda))\right]\Phi = 0, \tag{11}$$

where higher order terms in $O(1/n)$ are neglected, so that $\bar{L}_\alpha \approx \bar{L}_\alpha^{(0)}$ ($\alpha=1,2$). For $\lambda_r \approx 0$ with finite $\lambda_I$ by making use of the approximation

$$\Omega(\lambda) \approx \Omega_0 + \Omega_1\cos\lambda \xrightarrow{\lambda\to\lambda_r+i\lambda_I} \Omega_0 + \frac{\Omega_1}{2\cosh\lambda_I}(1+\cosh^2\lambda_I) - \frac{1}{2}\Omega_1\cosh\lambda_I(\lambda_r + i\tanh\lambda_I)^2. \tag{12}$$

Eq. (11) becomes

$$\frac{d^2\Phi}{d\lambda_r^2} + \frac{n^2}{\bar{L}_2}\left[\frac{\bar{L}_1^2}{4\bar{L}_2} + \Omega_0 + \frac{\Omega_1}{2\cosh\lambda_I}(1+\cosh^2\lambda_I) - \Omega - \frac{1}{2}\Omega_1\cosh\lambda_I(\lambda_r + i\tanh\lambda_I)^2\right]\Phi = 0. \tag{13}$$

When neglecting the weak $\lambda_r$-dependence in $\bar{L}_\alpha$ the solution of Eq. (13) is

$$\Phi(\lambda_r) = \exp\left[-n\sqrt{\frac{\Omega_1\cosh\lambda_I}{2\bar{L}_2}}(\lambda_r + i\tanh\lambda_I)^2/2\right] \tag{14}$$

with the eigenvalue

$$\Omega = \Omega_0 + \frac{\Omega_1}{2\cosh\lambda_I}\left(1+\cosh^2\lambda_I\right) + \frac{\bar{L}_1^2}{4\bar{L}_2}. \tag{15}$$

The full wave function in the second dimension is

$$\psi(\lambda_r) = \exp\left(-in\frac{\mathcal{F}}{2\bar{L}_2}\lambda_r\right)\exp\left[-n\sqrt{\frac{\Omega_1\cosh\lambda_I}{2\bar{L}_2}}\left(\lambda_r^2 - \tanh^2\lambda_I\right)/2\right], \tag{16}$$

where $\mathcal{F} \equiv \bar{L}_1 + \Delta(\lambda_I) \equiv \bar{L}_1 + \sqrt{2\bar{L}_2\Omega_1\cosh\lambda_I}\tanh\lambda_I$. For $\lambda_I = 0$ Eq. (16) reduces to Eq. (10), WABT-13, whereas WABT-92 is the solution of $\lambda_I$ for $\mathcal{F} = 0$. In contrast to WABT-92 the unified WABT does not require $\mathcal{F} = 0$. The rate of change of $\psi(\lambda_r)$ when divided by $n$ can be viewed to be composed of two parts; one from the term linear to $\lambda_r$, $\sim \mathcal{F}/2\bar{L}_2 \equiv \Xi_1/2$, the other from the term quadratic to $\lambda_r$, $\sim (\Omega_1\cosh\lambda_I/8\bar{L}_2)^{1/4}/\sqrt{n} \equiv \Xi_2$. In the large $n$ limit the condition $|\Xi_2| \gg 1/n$ yields faster variation of $\psi$ than $\chi$ in $\lambda$; while both $|\Xi_1| \ll 1$ and $|\Xi_2| \ll 1$ provide the asymptotic convergence. The condition $|\Xi_1| \ll 1$ addresses the complementariness of the two existing WABT theories. On one hand, the condition $\mathcal{F} = 0$ by WABT-92 seems too stringent--zero can be replaced by a small quantity $|\Xi_1| \ll 1$; on the other hand, the domain in parameter space can be enlarged by introducing finite $\lambda_I$, which is able to reduce $|\bar{L}_1|$ to $|\mathcal{F}| < |\bar{L}_1|$ when the equilibrium does not allow $\Xi \equiv \bar{L}_1/\bar{L}_2$ small enough for valid WABT-13.

Equations (15)-(16) with the condition $|\Xi_1| \ll 1$ are basic analytical results of the unified WABT.

## 3 The unified WABT applied to fluid ITG model

After having demonstrated the generality of the unified WABT approach, we will now apply it to a problem of great interest, the ITG mode, and use this example to demonstrate its effectiveness.

### 3.1 Basic equations of fluid ITG model in 2D Fourier-ballooning representation

The equation governing by the 2D high $n$ ITG mode (derived in a fluid model) is [39]

$$\left(\rho_s^2\nabla_\perp^2 - \frac{c_s^2}{\omega^2}\nabla_\parallel^2 - \frac{\omega - \hat{\omega}_{*e}}{\omega + \hat{\omega}_{*i}} - \frac{2\hat{\omega}_{de}}{\omega}\right)\varphi_n(r,\vartheta,\varsigma) = 0 \tag{17}$$

where $\nabla_\perp^2 \equiv \partial^2/\partial r^2 + (1/r^2)\partial^2/\partial\vartheta^2$, $\nabla_\parallel \equiv \mathbf{b}\cdot\nabla = (1/qR)[\partial/\partial\vartheta + q(r)\partial/\partial\varsigma]$, $c_s \equiv \sqrt{T_e(r)/m_i}$,

$$\hat{\omega}_{*e} \equiv \omega_{*e}(i/nq)\partial/\partial\vartheta \quad , \quad \omega_{*e} \equiv (k_\vartheta T_e(r)/eB)(d\ln n_0/dr) \quad , \quad \hat{\omega}_{*i} \equiv \omega_{*i}(i/nq)\partial/\partial\vartheta \quad ,$$

$$\omega_{*i} \equiv (k_\vartheta T_i(r)/eB)(d\ln p_i(r)/dr) \quad , \quad \hat{\omega}_{de} \equiv \omega_{de}(i/nq)(r\sin\vartheta\partial/\partial r + \cos\vartheta\partial/\partial\vartheta) \quad ,$$

$\omega_{de} \equiv -k_\vartheta T_e(r)/eBR$, $\rho_s \equiv \sqrt{m_i T_e(r)}/eB$, $k_\vartheta \equiv m/r_0$ and $q \equiv q(r_0)$. Here, $T_e(r)$ is the electron temperature, $n_0(r)$ is the plasma density in equilibrium, $T_i(r)$, $p_i(r)$ are the ion temperature and pressure respectively, $e$ is the unit charge, $B \approx B_0$ and $R \approx R_0$ are the magnetic field and major radius on the magnetic axis. Use is made of the linear $q(r)$ profile $q(r) := q + x/n$, and temperature profile $T_j(r) := T_j(r_0)(1 + x/k_\vartheta \hat{s} L_{T_j})$, but the quadratic density, pressure profile

$$\frac{d\ln n_0(r)}{dr} := \frac{1}{L_n} + \frac{x}{k_\vartheta \hat{s}}\left(\frac{\hat{s}_{L_n}}{R^2} - \frac{1}{L_n^2}\right), \quad \frac{d\ln p_i(r)}{dr} := \frac{1}{L_p} + \frac{x}{k_\vartheta \hat{s}}\left(\frac{\hat{s}_{L_n}}{R^2} - \frac{1}{L_n^2} - \frac{1}{L_{T_i}^2}\right), \tag{18}$$

where $L_n \equiv (d\ln n_0/dr)^{-1}_{r_0}$, $L_{T_j} \equiv (d\ln T_j/dr)^{-1}_{r_0}$, ($j = i, e$), $L_p \equiv (d\ln p_i/dr)^{-1}_{r_0}$, $\hat{s}_{L_n} \equiv (R^2/n_0)(d^2 n_0/dr^2)_{r_0}$ to derive

$$\nabla_\perp^2 = (k_\vartheta \hat{s})^2 \frac{\partial^2}{\partial x^2} + \frac{k_\vartheta^2}{(1 + x/nq\hat{s})^2}\left(\frac{1}{n^2 q^2}\frac{\partial^2}{\partial\vartheta^2}\right), \tag{19}$$

$$\nabla_\parallel \equiv \frac{1}{qR}\left[\frac{\partial}{\partial\vartheta} + q\left(1 + \frac{x}{nq}\right)\frac{\partial}{\partial\zeta}\right], \tag{20}$$

$$\hat{\omega}_{de} = \omega_{de}\left(1 + \frac{x}{k_\vartheta \hat{s} L_{T_e}}\right)\left[\hat{s}\sin\vartheta\left(i\frac{\partial}{\partial x}\right) + \frac{\cos\vartheta}{(1 + x/nq\hat{s})}\left(\frac{i}{nq}\frac{\partial}{\partial\vartheta}\right)\right], \tag{21}$$

$$\hat{\omega}_{*i} = \omega_{*i}\left(1 + \frac{x}{k_\vartheta \hat{s} L_{T_i}}\right)\left[1 + \frac{L_p x}{k_\vartheta \hat{s}}\left(\frac{\hat{s}_{L_n}}{R^2} - \frac{1}{L_n^2} - \frac{1}{L_{T_i}^2}\right)\right]\frac{1}{(1 + x/nq\hat{s})}\left(\frac{i}{nq}\frac{\partial}{\partial\vartheta}\right), \tag{22}$$

$$\hat{\omega}_{*e} = \omega_{*e}\left(1 + \frac{x}{k_\vartheta \hat{s} L_{T_e}}\right)\left[1 + \frac{L_n x}{k_\vartheta \hat{s}}\left(\frac{\hat{s}_{L_n}}{R^2} - \frac{1}{L_n^2}\right)\right]\frac{1}{(1 + x/nq\hat{s})}\left(\frac{i}{nq}\frac{\partial}{\partial\vartheta}\right), \tag{23}$$

where $\omega_{*e} \equiv k_\vartheta T_e(r_0)/eBL_n$, $\omega_{*i} \equiv k_\vartheta T_i(r_0)/eBL_p$, $\omega_{de} \equiv -k_\vartheta T_e(r_0)/eBR$ are henceforth redefined to be quantities on the rational surface.

Substituting Eqs. (18)-(23) into Eq. (17) yields the 2D high $n$ ITG mode in $(x, \vartheta)$ representation

$$\left\{ (k_\vartheta \rho_s \hat{s})^2 \frac{\partial^2}{\partial x^2} + \frac{(k_\vartheta \rho_s)^2}{(1+x/nq\hat{s})^2} \left(\frac{1}{nq}\frac{\partial}{\partial \vartheta}\right)^2 - \frac{\omega_s^2}{\omega^2}\left[\frac{\partial}{\partial \vartheta} + inq\left(1 + \frac{x}{nq}\right)\right]^2 - \frac{1}{(1+x/k_\vartheta \hat{s} L_{T_e})}\frac{\omega - \hat{\omega}_{*e}}{\omega + \hat{\omega}_{*i}} \right.$$
$$\left. - \frac{2\omega_{de}}{\omega}\left[\hat{s}\sin\vartheta\left(i\frac{\partial}{\partial x}\right) + \frac{\cos\vartheta}{(1+x/nq\hat{s})}\left(\frac{i}{nq}\frac{\partial}{\partial \vartheta}\right)\right] \right\} \varphi_n(x,\vartheta) = 0 \quad , \quad (24)$$

where $\omega_s \equiv c_s/qR$ and all quantities such as $\rho_s$, $\omega_{de}$, $c_s$, $q$ are defined on the rational surface, except for $\hat{\omega}_{*i}$ and $\hat{\omega}_{*e}$, which still are the operator as defined in Eqs. (22) and (23).

The 2D equation in $(x,l)$ representation follows from substituting

$$\varphi_n(x,\vartheta) = e^{-im\vartheta}\sum_l \varphi_l(x) e^{-il\vartheta} \quad (25)$$

into Eq. (24). The operator $\hat{\omega}_{*i}$ in the denominator as well as $\hat{\omega}_{*e}$ is treated by the Taylor expansion in $l/m \ll 1$. In contrast to the earlier work, e.g. Ref. [39], where only one TSB term was retained, the present analysis retains all TSB terms up to the second order, such as $(l/m)^2$, $x^2$. The following two rules are helpful in simplifying the derivation of the equation for $\varphi_l(x)$: (a) derivatives are applied only on the wave functions, not on the equilibrium quantities; and (b) the TSB term $x$ can be replaced by $l$, which is equivalent to the localization in $\lambda$, i.e., $\partial/\partial \lambda \gg \partial/\partial k$. The equation for $\varphi_l(x)$ is

$$\left[k_\vartheta^2 \rho_s^2 \hat{s}^2 \frac{d^2}{dx^2} + \frac{\omega_s^2}{\omega^2}(x-l)^2 - k_\vartheta^2 \rho_s^2 - \frac{\omega - \omega_{*e}}{\omega + \omega_{*i}}\right]\varphi_l(x) - \frac{\omega_{de}}{\omega}\left[\left(1 + \hat{s}\frac{d}{dx}\right)\varphi_{l+1}(x) + \left(1 - \hat{s}\frac{d}{dx}\right)\varphi_{l-1}(x)\right]$$
$$+ \left\{\left[-2k_\vartheta^2 \rho_s^2\left(1 - \frac{1}{\hat{s}}\right) + C\right]\varphi_l(x) - \frac{\omega_{de}}{\omega}\left(1 - \frac{1}{\hat{s}}\right)[\varphi_{l+1}(x) + \varphi_{l-1}(x)]\right\}\left(\frac{l}{m}\right) \quad (26)$$
$$+ \left\{\left[-k_\vartheta^2 \rho_s^2\left(1 - \frac{4}{\hat{s}} + \frac{3}{\hat{s}^2}\right) - D\right]\varphi_l(x) + \frac{\omega_{de}}{\omega}\frac{1}{\hat{s}}\left(1 - \frac{1}{\hat{s}}\right)[\varphi_{l+1}(x) + \varphi_{l-1}(x)]\right\}\left(\frac{l}{m}\right)^2 = 0$$

with

$$A_n \equiv \frac{1}{\hat{s}}\left(1 - \frac{r_0 L_n}{R^2}\hat{s}_{L_n} + \frac{r_0}{L_n} - \frac{r_0}{L_{T_e}}\right), \quad A_p \equiv \frac{1}{\hat{s}}\left(1 - \frac{r_0 L_p}{R^2}\hat{s}_{L_n} + \frac{r_0 L_p}{L_n^2} + \frac{r_0 L_p}{L_{T_i}^2} - \frac{r_0}{L_{T_i}}\right), \quad (27)$$

$$B_n \equiv \frac{1}{\hat{s}^2}\left(1 - \frac{r_0 L_n}{R^2}\hat{s}_{L_n} + \frac{r_0}{L_n}\right)\left(1 - \frac{r_0}{L_{T_e}}\right), \quad B_p \equiv \frac{1}{\hat{s}^2}\left(1 - \frac{r_0 L_p}{R^2}\hat{s}_{L_n} + \frac{r_0 L_p}{L_n^2} + \frac{r_0 L_p}{L_{T_i}^2}\right)\left(1 - \frac{r_0}{L_{T_i}}\right), \quad (28)$$

$$C \equiv \frac{\omega - \omega_{*e}}{\omega + \omega_{*i}}\frac{\omega_{*i}}{\omega + \omega_{*i}}(1 - A_p) + \frac{\omega_{*e}}{\omega + \omega_{*i}}(1 - A_n) + \frac{\omega - \omega_{*e}}{\omega + \omega_{*i}}\frac{r_0}{\hat{s}L_{T_e}}, \quad (29)$$

$$D \equiv \left(C - \frac{\omega - \omega_{*e}}{\omega + \omega_{*i}}\frac{r_0}{\hat{s}L_{T_e}}\right)\left[\frac{\omega_{*i}}{\omega + \omega_{*i}}(1 - A_p) + \frac{r_0}{\hat{s}L_{T_e}}\right] + \frac{\omega - \omega_{*e}}{\omega + \omega_{*i}}\frac{\omega_{*i}}{\omega + \omega_{*i}}(A_p - B_p) + \frac{\omega_{*e}}{\omega + \omega_{*i}}(A_n - B_n) + \frac{\omega - \omega_{*e}}{\omega + \omega_{*i}}\left(\frac{r_0}{\hat{s}L_{T_e}}\right)^2$$
$$(30)$$

The set of Eqs. (26)-(30) will also be numerically solved (by finite difference method) in section 4 with the natural boundary condition provided by the results obtained in this section from WABT.

Applying ballooning transformation Eq. (1) to Eq. (26) yields the 2D ITG equation in the form of Eq. (2). The ballooning operator and the global eigen-value defined in Eq. (2) are

$$L_0(k,\lambda) \equiv \frac{1}{\eta^2}\frac{\partial^2}{\partial k^2} + \eta^2 k^2 - \frac{2q}{\hat{s}}\left[\cos(k+\lambda) + \hat{s}k\sin(k+\lambda)\right], \quad (31)$$

$$\Omega \equiv -(\eta/\hat{s})^2 - (\eta/k_g\rho_s\hat{s})^2(\omega-\omega_{*e})/(\omega+\omega_{*i}) \quad (32)$$

where $\eta^2 \equiv \omega k_g \rho_s \hat{s}/\omega_s$. The next two coefficients of Eq. (2), $L_\alpha(k,\lambda)$ ($\alpha=1,2$), are

$$L_1(k,\lambda) \equiv \frac{1}{q}\left[-\frac{2\eta^2}{\hat{s}^2}\left(1-\frac{1}{\hat{s}}\right) + \frac{C\eta^2}{(k_g\rho_s\hat{s})^2} + \frac{2q}{\hat{s}}\left(1-\frac{1}{\hat{s}}\right)\cos(k+\lambda)\right], \quad (33)$$

$$L_2(k,\lambda) \equiv \frac{1}{q^2}\left[-\frac{\eta^2}{\hat{s}^2}\left(1-\frac{4}{\hat{s}}+\frac{3}{\hat{s}^2}\right) - \frac{D\eta^2}{(k_g\rho_s\hat{s})^2} - \frac{2q}{\hat{s}^2}\left(1-\frac{1}{\hat{s}}\right)\cos(k+\lambda)\right]. \quad (34)$$

Their simple form in $k$-dependence of $L_\alpha(k,\lambda)$ makes it possible to obtain $\bar{L}_\alpha$ as the ballooning wave function is modeled by

$$\chi(k,\lambda) = e^{i\eta^2(k-k_*(\lambda))^2/2} \quad (35)$$

(see Eq. (B3) in Ref. [39]), where $k_*$ is the minimum of the potential well of the ballooning equation determined by

$$k_* = (q/\hat{s}\eta^2)\left[(\hat{s}-1)\sin(k_*+\lambda) + k_*\hat{s}\cos(k_*+\lambda)\right], \quad (36)$$

$$\bar{L}_1(\lambda) = \frac{1}{q}\left[-\frac{2\eta^2}{\hat{s}^2}\left(1-\frac{1}{\hat{s}}\right) + \frac{\eta^2}{(k_g\rho_s\hat{s})^2}C + \frac{2q}{\hat{s}}\left(1-\frac{1}{\hat{s}}\right)I_c\right], \quad (37)$$

$$\bar{L}_2(\lambda) = \frac{1}{q^2}\left[-\frac{\eta^2}{\hat{s}^2}\left(1-\frac{4}{\hat{s}}+\frac{3}{\hat{s}^2}\right) - \frac{\eta^2}{(k_g\rho_s\hat{s})^2}D - \frac{2q}{\hat{s}^2}\left(1-\frac{1}{\hat{s}}\right)I_c\right] \quad (38)$$

with

$$I_c \equiv e^{-1/(4\operatorname{Im}\eta^2)}\cos\left[\operatorname{Re}k_*(\lambda) + \operatorname{Re}\eta^2\operatorname{Im}k_*(\lambda)/\operatorname{Im}\eta^2 + \lambda\right]. \quad (39)$$

For a quality check, these analytical expressions Eqs. (36)-(39) ought to be sample-verified *a posteriori* by direct numerical integration using $\chi(k,\lambda)$ obtained from the shooting solution.

### 3.2 Numerical solution of 2D ballooning ITG mode by iterative shooting code

An iterative shooting code ITG-WABT has been developed to solve the $\lambda$ differential Eq. (11) associated with the ITG ballooning equation

$$\left\{\frac{1}{\eta^2}\frac{\partial^2}{\partial k^2}+\eta^2 k^2-\frac{2q}{\hat{s}}\left[\cos(k+\lambda)+\hat{s}k\sin(k+\lambda)\right]-\Omega(\lambda)\right\}\chi(k,\lambda)=0 \tag{40}$$

with $\eta^2 \equiv \omega k_g \rho_s \hat{s}/\omega_s$ and the global eigen-value expressed by Eq. (32).

The operation of the shooting code starts with the initial guess of the mode frequency $\omega \to \omega^{(0)}$ corresponding to the initial guess of the global eigen-value $\Omega^{(0)}$ per Eq. (32). Use is made of the initial guess $\omega^{(0)}$ to solve the ballooning equation Eq. (31) to obtain the local eigen-value $\Omega^{(0)}(\lambda;\omega^{(0)})$ and the ballooning wave function $\chi^{(0)}(k,\lambda;\omega^{(0)})$, then to calculate $\bar{L}_{1,(0)}(\lambda;\omega^{(0)})$, $\bar{L}_{2,(0)}(\lambda;\omega^{(0)})$ via Eqs. (37)-(39) for a given prescribed $\lambda_I$. Substituting the three quantities $\Omega^{(0)}(\lambda;\omega^{(0)})$, $\bar{L}_{1,(0)}(\lambda;\omega^{(0)})$ and $\bar{L}_{2,(0)}(\lambda;\omega^{(0)})$ into Eq. (11) results in the wave function of the second dimension $\Phi^{(0)}(\lambda)$ and the global eigen-value $\Omega^{(1)}$, which yields the first iterative mode frequency $\omega^{(1)}$ by Eq. (32). The same procedure is repeated until $\left|1-\omega^{(i)}/\omega^{(i+1)}\right|<10^{-4}$ is satisfied. The sample numerical check on the analytical model Eqs. (36)-(39) for $\bar{L}_\alpha$ are carried out as a component of the code.

The following parameters (relevant to JET) are chosen to explore the eigen-mode $\Phi(\lambda)$ localized at $\lambda \to 0$: $R=2.96\text{m}$, $a=1.25\text{m}$, $T_e=T_i=3\text{keV}$, $B=3\text{T}$. It is assumed that $k_g\rho_s=0.31$, $\hat{s}=1.5$, $q=3.5$, $L_n/R=-0.02$, $\hat{s}_{L_n}=0$, and $L_{T_i}=L_{T_e}$. The toroidal mode number was chosen to be $n=40$. This set of parameters mostly satisfies the condition $|\Xi|<1$ while keeping the value of $\left|\Xi\bar{L}_1\right|/4$ moderate.

In Fig. 1, we plot the normalized global eigen-frequency $\omega/|\omega_{*e}|$ as a function of $\bar{\eta}_i \equiv |\omega_{*i}/\omega_{*e}|$ for the ITG mode predicted by HBT (I) and VBT (II) as calculated in Ref. [43], and the new WABT with $\lambda_I=0$ (III) respectively. The solid lines are growth rates, and the dotted-dashed lines are real frequencies for (I)-(III). As shown in Fig. 1 WABT are in qualitative agreement with those of HBT demonstrating that the linear growth rates predicted by the CBT (HBT) could be qualitatively valid even though the solvability condition is not satisfied. In particular, CBT's prediction is much closer to WABT's than VBT's. The growth rates of HBT and WABT are larger than the growth rates of VBT, because the localization of the VBT mode away from the region of bad curvature reduces the growth rates. Modes with the VBT structure are then likely to capture the physics of anomalous transport only for not small $\Xi_1$ region. We note that the growth rate

predicted by HBT is not always greater than that by WABT, even though bad curvature favors the former more than the latter. This is an indication of the exceeding contribution from second dimension and the non-trivial effects from TSB terms.

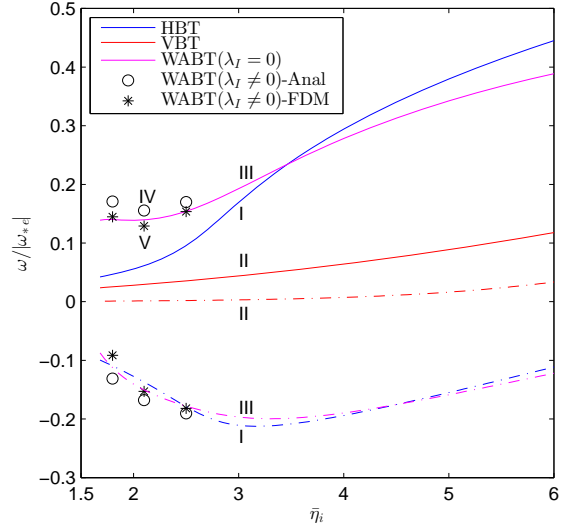

Fig. 1 Normalized eigen-frequency of 2D ITG mode versus $\bar{\eta}_i$. The solid line corresponds to the growth rate and the dotted-dashed line to the real frequency for parameters $k_g \rho_s = 0.31$, $\hat{s} = 1.5$, $q = 3.5$, $L_n / R = -0.02$, $\hat{s}_{L_n} = 0$, $L_{T_i} = L_{T_e}$, and $n = 40$. The Roman letters I, II, III represent the HBT, VBT, WABT with $\lambda_I = 0$; IV, (V) represent WABT with $\lambda_I \neq 0$ analytic (finite difference method) respectively.

For $\bar{\eta}_i \geq 2.5$ it is found that $|\Xi| \leq 0.23$, and the attempt to make $|\Xi_1| < |\Xi|$ by introducing finite $\lambda_I$ does not seem to be significant to this purpose. For lower $\bar{\eta}_i$, $|\Xi|$ is found to increase rapidly, and finite $\lambda_I$ would be helpful to save WABT in this parameter domain. For example, at $\bar{\eta}_i = 2.1$, $\lambda_I = -0.9$ (for minimum $|\Xi_1|$) reduces the second small parameter to $|\Xi_1| \approx 0.14$ compared with $|\Xi| \approx 0.37$. Three points are illustrated in Fig.1 for $\bar{\eta}_i = 1.8$, 2.1, 2.5 respectively. The analytical eigenvalues (hollow circles as marked by IV) are calculated by making use of Eq. (15), where $\lambda_I$ are taken at the minimum $|\Xi_1|$. The numerical verification on the approximations of Eq. (15) have been performed to endorse the validity. Corresponding to the analytical solutions labeled by three hollow circles, are the finite difference solutions denoted by three stars (V), as explained in the section 4.

The (complex) parameter $\Xi$ ($\Xi_1$) introduced in WABT appears to be the hidden parameter in the ballooning equation in the sense that it is a synthesized parameter of the "second dimension". For $|\Xi| > 1$ the mode may appear to be fully up-down asymmetric but has a lower growth rate. In the opposite limit for $|\Xi| < 1$, it acquires weak up-down asymmetry and has a higher growth rate.

Surely, we have no way to rule out the possibility of $|\Xi| \approx 1$ for parameters of physical interest. However, it may not be so demanding or even necessary to develop an independent theory for the marginal case because the ballooning theory, as an asymptotic theory, is plausible only for limited parameter regions. One of the parameters is well-known, namely the inverse of toroidal mode number $1/n$. The second small parameter $\Xi$, essentially, measures the ballooning orientation (horizontal for $|\Xi|<1$ vs vertical for $|\Xi|>1$) for modes with poloidal localization. As long as one is able to solve in either segment of $\Xi$, it would be possible step by step to approach $|\Xi| \approx 1$ from either end numerically using the initial guess for local modes. Such a procedure is partially demonstrated in the next section of the paper for $|\Xi|<1$ by taking WABT to be the natural boundary condition to solve the system (26)-(30).

## 3.3 Analytical mode structure of 2D ITG in real space

The analytical expression of the 2D wave function is constructed on the basis of the factorized form $\varphi(k,\lambda) = \psi(\lambda)\chi(k,\lambda)$ with $\psi(\lambda)$ given by Eq. (16), $\chi(k,\lambda)$ by Eq. (35), associated with the $\lambda$-dependent parameters $k_*(\lambda)$, $\bar{L}_1(\lambda)$, $\bar{L}_2(\lambda)$, and other mode frequency dependent parameters $\eta^2$, $\Omega_1$.

Use is made of Eq. (1) to obtain the wave function in $(x,l)$ representation

$$\varphi_l(x) = \frac{1}{2\pi}\int_{-\pi}^{\pi} d\lambda_r G[x-l;\lambda] \exp\left(-in\frac{\mathcal{F}}{2\bar{L}_2}\lambda_r - i\lambda_r l - n\sqrt{\frac{\Omega_1 \cosh\lambda_I}{2\bar{L}_2}}(\lambda_r^2 - \tanh^2\lambda_I)/2\right) \quad (41)$$

with the first integral over $k$ given by

$$G[x-l;\lambda] \equiv \int_{-\infty}^{+\infty} dk e^{ik(x-l)} e^{i\eta^2(k-k_*(\lambda))^2/2} = \sqrt{\frac{2i\pi}{\eta^2}} e^{ik_*(\lambda)(x-l) - \frac{i}{2\eta^2}(x-l)^2} . \quad (42)$$

The integration over $\lambda_r$ is performed by the standard steepest descent method. The saddle point is $\lambda_{r0} \equiv -i\sqrt{2\bar{L}_2/\Omega_1 \cosh\lambda_I}\left(\mathcal{F}/2\bar{L}_2 + l/n\right)$ where $\lambda$-dependent functions in the integrand such as $\bar{L}_2(\lambda)$, $\mathcal{F}(\lambda)$, $k_*(\lambda)$ are set to be the value at $\tilde{\lambda}_{r0} \equiv \lambda_{r0} + i\lambda_I$ in a perturbative manner for $\lambda_{r0}$. The integration contour on the complex $\lambda_r$ plane around the saddle point $\lambda_{r0}$ is determined by the angle $\theta_0 \equiv \text{Arg}\left(-\sqrt{\Omega_1 \cosh\lambda_I/8\bar{L}_2}\right)$ to real axis. This integration results in

$$\varphi_l(x) \approx \sqrt{\frac{i}{\eta^2 n\mathcal{A}}} e^{ik_*(\tilde{\lambda}_{r0})(x-l) - \frac{i}{2\eta^2}(x-l)^2} e^{n\sqrt{\frac{\Omega_1\cosh\lambda_I}{2\bar{L}_2}}\tanh^2\lambda_I/2 + i\theta_0} e^{-\frac{n}{2}\sqrt{\frac{2\bar{L}_2}{\Omega_1\cosh\lambda_I}}\left(\frac{\mathcal{F}}{2\bar{L}_2}+\frac{l}{n}\right)^2} , \quad (43)$$

where $\mathcal{A} \equiv \left|\sqrt{\Omega_1 \cosh \lambda_I / 2\bar{L}_2}\right|$.

Substituting Eq. (43) into Eq. (25) we obtain the analytical expression of wave function in the $(x,\vartheta)$ representation

$$\varphi(x,\vartheta) = \sqrt{\frac{i}{\eta^2 n \mathcal{A}}} e^{-im\vartheta} e^{\frac{n}{2}\sqrt{\frac{\Omega_1 \cosh \lambda_I}{2\bar{L}_2}}\left(\tanh^2 \lambda_I - \frac{\mathcal{F}^2}{2\bar{L}_2 \Omega_1 \cosh \lambda_I}\right) + i\theta_0}$$
$$\sum_l \exp\left[ik_*(\tilde{\lambda}_{r0})(x-l) - \frac{i(x-l)^2}{2\eta^2}\right] e^{-\sqrt{\frac{2\bar{L}_2}{\Omega_1 \cosh \lambda_I}}\frac{l^2}{2n} - il\left(\vartheta - i\frac{\mathcal{F}}{2\bar{L}_2}\sqrt{\frac{2\bar{L}_2}{\Omega_1 \cosh \lambda_I}}\right)}. \quad (44)$$

The up-down asymmetric ballooning orientation is described by the angle

$$\vartheta_* \equiv -\text{Im}\left(\frac{\mathcal{F}}{2\bar{L}_2}\sqrt{\frac{2\bar{L}_2}{\Omega_1 \cosh \lambda_I}}\right). \quad (45)$$

## 4 WABT as the natural boundary condition for numerical solution of 2D high *n* local eigenmodes

In the sheared slab model, the natural boundary condition for 1D high *n* radially local eigenmodes is imposed to be the outgoing waves on the ground state [44]. However, for 2D high *n* local eigenmodes, the suitable boundary condition is not simply the same radial boundary condition plus the periodicity in the poloidal direction. Such a straightforward but naïve idea would lead to a first order partial differential equation allowing infinite number of solution; and with no physical insight to guide, one would be hard pressed to settle on which one among these asymptotic solutions might yield the most unstable mode. In the context of the ballooning concept, however, the lowest order is constructed along the field line; a close correlation between $x$ and $\vartheta$, thus, has been embedded in a certain manner to recover the most unstable modes. For example, the so-called flux tube representation [45] is more appropriate for numerical simulation oriented to the eigenmode problem [46], even though it is *not* a full 2D representation. Another interesting approach to set up the boundary condition for local eigenmode is to convert the numerical results from the initial value code into the Fourier-ballooning representation [41,47].

Alternatively, a natural boundary condition for local 2D eigenmodes can be provided by the WABT structure as developed in the previous section; remember that WABT structure is authentic 2D. This route is also the beginning of a numerical verification of the theory. In this section, the finite difference code ITG-DNBC will be used to solve the 2D eigen-value problem directly by taking WABT to be the mathematical boundary condition at sufficiently large $|x| \geq \sqrt{n}$; the numerical result in the entire region are also compared with the prediction of WABT.

Use is made of the standard finite difference method to divide the $(x,l)$ rectangular domain into $(2n_x+1)\times(2n_l+1)$ grids with the wave functions on the nodes labeled by $\varphi_l^j \equiv \varphi_l(x_j)$, where $j\in[-n_x,n_x]$, $l\in[-n_l,n_l]$. The Dirichlet boundary condition $\varphi_{\pm n_l}^j = 0$ is introduced for the cut-off at large $|l|$ end, whereas the natural boundary condition $d\ln\varphi_l(x)/dx = b_l$ is applied radially at the neighborhood of $j = \pm n_x$ to give

$$\varphi_l^{\pm n_x} = \frac{2 \pm b_l^{\pm n_x}\Delta x}{2 \mp b_l^{\pm n_x}\Delta x}\varphi_l^{\pm n_x \mp 1}, \qquad (46)$$

where $b_l^{\pm n_x}$ is computed using the solution of WABT. It has been numerically sample-verified to use the analytical form of mode structure Eq. (43), from which the logarithmic derivative of wave functions is

$$\frac{d}{dx}\ln\varphi_l(x) = i\left(k_* - \frac{x-l}{\eta^2}\right), \qquad (47)$$

where $k_* = k_*(\tilde{\lambda}_{r0})$ and $\eta^2$ is associated with the eigen-frequency of the numerical WABT. The discretization of Eq. (26) with the boundary conditions Eq. (47) yields the matrix equation

$$\mathbf{M}\boldsymbol{\Phi} = \Omega\boldsymbol{\Phi}, \qquad (48)$$

where $\boldsymbol{\Phi} = \left(\boldsymbol{\Phi}^{-n_x+1}, \boldsymbol{\Phi}^{-n_x+2}, \cdots, \boldsymbol{\Phi}^{n_x-2}, \boldsymbol{\Phi}^{n_x-1}\right)^T$, $\boldsymbol{\Phi}^j = \left(\varphi_{-n_l+1}^j, \varphi_{-n_l+2}^j, \cdots, \varphi_{n_l-2}^j, \varphi_{n_l-1}^j\right)^T$, and $\mathbf{M}$ is an asymmetric positive definite matrix. The dimension of $\mathbf{M}$ is $(2n_x-1)(2n_l-1)\times(2n_x-1)(2n_l-1)$.

Due to the dependence of the matrix $\mathbf{M}$ on the eigen-frequency $\omega$, Eq. (48) needs to be solved numerically using an iterative method. The global eigen-value of WABT, denoted by $\Omega^{(0)}$, is taken to be the initial guess with corresponding mode frequency $\omega^{(0)}$; they are then substituted into the coefficient matrix $\mathbf{M}^{(0)}$. The shifted inverse power method [48] is adopted to solve Eq. (48). We first subtract $\Omega^{(0)}$ from both sides of Eq. (48) to obtain

$$\left(\mathbf{M}^{(0)} - \Omega^{(0)}\right)\boldsymbol{\Phi} = \left(\Omega - \Omega^{(0)}\right)\boldsymbol{\Phi}. \qquad (49)$$

Then, we attempt to obtain the eigen-value of Eq. (49) as close as possible to zero from Eq. (49), i.e. the closest one of Eq. (48) to $\Omega_0$, by using the inverse power method. The concrete procedure is briefly described as follows. ① The 2-D wave functions of Eq. (43) obtained from WABT, $\boldsymbol{\Phi}^{(0)}$, are taken as the guess of Eq. (49); ② We adopt the LU decomposition technique (any non-singular matrix can be written as a product of a lower triangular and an upper triangular matrix) to solve

$\left( \mathbf{M}^{(0)} - \Omega^{(0)} \right) \mathbf{\Phi}^{(1)} = \mathbf{\Phi}^{(0)}$ for $\mathbf{\Phi}^{(1)}$; ③ $\mathbf{\Phi}^{(1)}$ is normalized to $\sqrt{\left( \mathbf{\Phi}^{(1)} \right)^* \mathbf{\Phi}^{(1)}}$, where $\left( \mathbf{\Phi}^{(1)} \right)^*$ is the complex conjugate of $\mathbf{\Phi}^{(1)}$; ④ Multiplying $\left( \mathbf{\Phi}^{(1)} \right)^*$ on both sides of Eq. (49) yields $\Delta \Omega^{(1)} = \left( \mathbf{\Phi}^{(1)} \right)^* \left( \mathbf{M}^{(0)} - \Omega^{(0)} \right) \mathbf{\Phi}^{(1)}$; ⑤ Repeat the steps ②-④ with $\mathbf{\Phi}^{(0)}$ replaced by $\mathbf{\Phi}^{(1)}$ till a sufficiently small value of $\left| 1 - \Delta \Omega^{(k+1)} / \Delta \Omega^{(k)} \right|$ is reached. Now, the obtained eigen-pairs, $\mathbf{\Phi}^{(k)}$ and $\Delta \Omega^{(k)}$, are the eigen-function and eigen-value closest to zero of Eq. (49). The eigen-value of Eq. (48) is, thus, approximated by $\Omega^{(1)} = \Omega^{(0)} + \Delta \Omega^{(k)}$ with corresponding eigen-frequency $\omega^{(1)}$, which is employed to calculate the coefficient matrix $\mathbf{M}^{(1)}$. Then, we substitute $\mathbf{M}^{(1)}$ and $\Omega^{(1)}$ into Eq. (49) to replace $\mathbf{M}^{(0)}$ and $\Omega^{(0)}$, respectively. Eq. (49) is in turn to be solved again for further correction to eigen-value till the convergence condition $\left| 1 - \Omega^{(i+1)} / \Omega^{(i)} \right| < \varepsilon$ is satisfied.

The comparison of numerical results obtained from the finite difference method with analytical results of WABT is shown below in Figs. 2 and 3 for parameters the same as those in Fig. 1, but for $\bar{\eta}_i = 2.5$ and $\lambda_I = 0$. The finite difference mesh is chosen to be $x \in [-20, 20]$, $l \in [-10, 10]$, $n_l = 10$, $n_x = 40$. The eigen-frequencies obtained by finite difference method and WABT are, respectively, $\omega / |\omega_{*e}| = -0.181 + 0.154i$ and $\omega / |\omega_{*e}| = -0.178 + 0.154i$. The difference is $1.3\%$, consistent with the ballooning error in the order of $1/n$. The second small parameter $\Xi$ for this parameter set is found to be 0.23.

The two sets of wave functions for the central Fourier mode $\varphi_0(x)$ are displayed in Fig. 2, where Fig. 2(b) is obtained from Eq. (43) for $\lambda_I = 0$ according to WABT in the entire region, and Fig. 2(a) is the result of the 2D numerical calculation by the finite difference method with a boundary condition consistent with Eq. (43) for $\lambda_I = 0$. The results indicate good agreements between analytical solution given by the WABT and the 2D numerical solution.

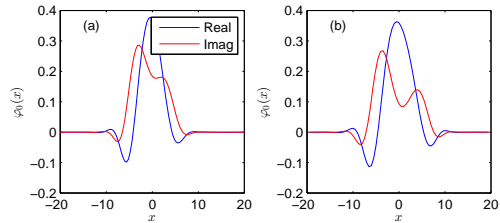

Fig. 2 1-D wave function $\varphi_0(x)$ given by (a) finite difference method and (b) WABT for same parameters as in Fig. 1, but for $\bar{\eta}_i = 2.5$ and $\lambda_I = 0$.

The 2D mode structure is shown in Fig. 3 via the contour plot of $\text{Re}[\varphi_n(r,\vartheta,0)]$. Figure 3(a) is the result by the 2D numerical calculation, and Fig. 3(b) is obtained from Eq. (43) for $\lambda_I = 0$ and Eq. (25) according to the WABT. The radial position is determined by the mapping $(1+x/m\hat{s})(r_0/a)$. The weak up-down asymmetry of the mode structure is illustrated through the measure $\vartheta_* \equiv -\text{Im}\left[(\bar{L}_1/\bar{L}_2)\sqrt{(\bar{L}_2/2\Omega_1)}\right]\bigg|_{\lambda=\tilde{\lambda}_{r0}} = 0.22$ ($\sim 13°$). This kind of the asymmetry can provide the torque for toroidal rotation similar to the situations discussed in Ref. [39].

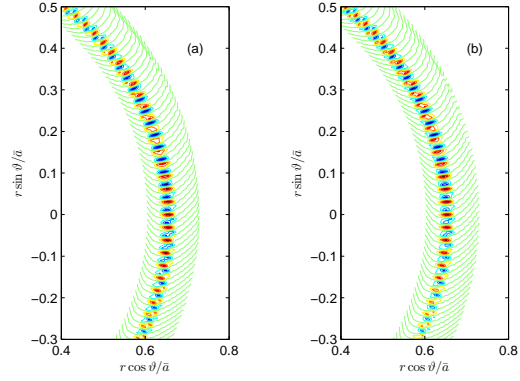

Fig. 3 Level contour of $\text{Re}[\varphi_n(r,\vartheta,0)]$ in poloidal cross section obtained from (a) finite difference method and (b) WABT with parameters same as in Fig. 2.

## 5  Concluding remarks

On basis of the two existing theories WABT-92 [11] and WABT-13 [43], the unified ballooning theory with weak up-down asymmetric mode structure (WABT) is presented. Major physical results are illustrated and obtained by studying a specific model: the fluid formulation of ITG. It is shown that the weak up-down asymmetry is the manifestation of the non-trivial effects from the translational symmetry breaking (TSB) terms beyond the ballooning equation. All TSB terms are included in the computation to assure quantitative validity. For up-down symmetric equilibrium the ballooning equation is CP conservative; in the Fourier-ballooning representation, this translates as the invariance under the combined reflection $(k,\lambda) \to (-k,-\lambda)$ (equivalent to $(\eta,\eta_0) \to (-\eta,-\eta_0)$ in the Connor-Hastie-Taylor representation [5,6]). The CP conservation, within the framework of ballooning equation, limits the poloidal localization to be around either 0 or $\pi$. When TSB terms are taken into account, CP conservation is broken, and weak up-down asymmetry follows. Such an asymmetry has also been theoretically observed in Connor-Hastie-Taylor representation for *up-down asymmetric equilibrium*[49].

We believe that of all the structures that have been investigated, WABT structure may be the most relevant to anomalous transport; WABT assures the asymptotic convergence in a large parameter region as compared to CBT (HBT), while it comes with a growth rate higher than VBT. In fact, in certain parameter regimes, WABT has a growth rate even higher than CBT (HBT). An interesting message: to improve the tokamak confinement, one may want to choose parameters so that WABT is hard to excite by tuning up the equilibrium to minimize the range of validity for WABT, in particular, away from heating region.

The analytical mode structure of WABT derived in the fluid ITG model is found to be vastly useful. It first serves as a natural boundary condition for the 2D mathematical local eigenmode problem in the real space, *i.e,* $(x,\vartheta)$ or $(x,l)$ representation. In the region of validity (sufficiently small $1/\sqrt{n}$ and $|\Xi|$), the mathematical results can be compared with WABT's throughout the entire local radial domain, *i.e,* not limited to $|x| \geq \sqrt{n}$. The comparison provides the numerical verification of WABT as shown in section 4 by making use of the finite difference method to solve the mathematical problem. The good agreements suggest that the analytical mode structure of WABT could be fully trusted for calculating quasi-linear quantities of interest to anomalous transport, *e.g.*, poloidal and toroidal Reynolds stress similar to those done in VBT [25,39], moment fluxes and other wave-induced physical quantities. Equally relevant is the realization that the method can be extended to the ballooning-like modes near plasma edge, taking asymptotic WABT solution to be the inner boundary condition, while leaving the outer boundary condition to be determined by the physical environment such as a conducting wall.

## Acknowledgement

This research was supported by CAS Program for Interdisciplinary Collaboration Team, by the JSPS-NRF-NSFC A3 Foresight Program in the field of Plasma Physics (NSFC-11261140328), by ITER-China Program (2014GB124005), by the Fundamental Research Funds for the Central Universities (WK2030040052), and by the U.S. Dept. of Energy Grant DE-FG02-04ER-54742.